\documentclass[shortnote,twocolumn]{jpsj3}
\usepackage{txfonts}
\usepackage{color}
\usepackage{amsmath}

\title{Large Single Crystal Growth and Transport Properties of KCd$_4$As$_3$ and RbCd$_4$As$_3$}

\author{Keigo Ono, Yosuke Goto, Masayuki Murata, Hiroyasu Matsuura, and Chul-Ho Lee\thanks{c.lee@aist.go.jp}}
\inst{National Institute of Advanced Industrial Science and Technology (AIST), Tsukuba, Ibaraki 305-8568, Japan} 

\abst{
Large-size KCd$_4$As$_3$ and RbCd$_4$As$_3$ single crystals were grown using the flux method. 
Appropriate flux for growing single crystals was determined to be Cd$_2$As, resulting in the increase in their size to be a centimeter for KCd$_4$As$_3$.
The KCd$_4$As$_3$ and RbCd$_4$As$_3$ crystals exhibited metal-like temperature dependences of their electrical resistivities with negative Seebeck coefficients, characterizing them as n-type thermoelectric materials.
}

\begin{document}
\maketitle
143-Zintl phase compounds $AM_4X_3$ ($A$: alkali metal, $M$: transition metal, $X$: pnictogen) have recently been identified as a new class of high-performance thermoelectric materials\cite{NaZn4As3_aichi, RbZn4As3_Ono, RbCd4As3_Ono, RbZn4P3calc_2022, CsZn4As3calc_2024, CsMn4As3calc_2025, CsCd4As3calc_2026}. 
The compounds possess conflicting properties of low electrical resistivity ($\rho$), high Seebeck coefficient ($S$), and low thermal conductivity ($\kappa$)
resulting in the dimensionless figure-of-merit ($ZT = S^{2}T/\rho\kappa$) of 0.53 and 0.43 in p-type RbZn$_4$As$_3$ and n-type RbCd$_4$As$_3$, respectively\cite{RbZn4As3_Ono, RbCd4As3_Ono}.
The origin of these properties is worth investigating to establish a strategy for developing new high-performance thermoelectric materials.
Intriguingly, the 143-Zintl phases also present extremely high Hall mobilities ($>$ 10$^3$ cm$^2$/Vs for electron carriers at room temperature) despite their low lattice thermal conductivity.\cite{RbCd4As3_Ono,NaCd4As3_topo,NaCd4As3_Hall}
An hypothesis has been proposed that asymmetric Hall mobility between the electron and hole carriers, 
associated with a narrow energy band gap, contributes to the high power factor ($PF = S^{2}/\rho$) of these phase\cite{RbCd4As3_Ono}. 
The 143-Zintl phases have also attracted attention for their topological properties.\cite{NaCd4As3_topo, RbCd4As3_calc, NaCd4As3_Hall, NaCd4As3_2025, NaZn4As3_topo}
For example, NaCd$_4$As$_3$ has been proposed as a topological insulator and first-principles calculations have claimed topological band inversion in RbCd$_4$As$_3$.\cite{NaCd4As3_topo, NaCd4As3_Hall, NaCd4As3_2025, RbCd4As3_calc}.
Thus, by investigating the 143-Zintl phases, we could understand both the origin of high-performance thermoelectric properties and the nature of topological materials.
However, large-sized, high-quality single crystals are essential for clarifying the electronic structures of these phases.
Thus far, NaZn$_4$As$_3$ and NaCd$_4$As$_3$ single crystals have been grown to millimeter-order size under Zn and Cd fluxes, respectively.\cite{NaCd4As3_topo, NaCd4As3_Hall, NaCd4As3_2025, NaZn4As3_topo}
In this study, we demonstrate that single crystals of $A$Cd$_4$As$_3$ can be grown to centimeter-order size under a Cd$_2$As flux.


Single crystals of KCd$_4$As$_3$ and RbCd$_4$As$_3$ crystallized in the rhombohedral structure (space group $R\bar{3} m$) 
were grown under a Cd or Cd$_2$As self-flux. 
To synthesize the KAs, RbAs, and Cd$_2$As precursors, K (3N, chunk), Rb (3N+, chunk), Cd (6N, shot), and As (7N5, chunk) were mixed at a stoichiometric composition, 
loaded into alumina crucibles, and sealed in stainless-steel containers filled with Ar gas. \cite{Kihou_KFe2As2}
The containers of KAs, RbAs, and Cd$_2$As were then heated to 650$^\circ$C for 10 h, 650$^\circ$C for 20 h, and 750$^\circ$C for 20 h, respectively. 
The precursors were mixed with Cd and As at the compositions described below, loaded into alumina crucibles, and sealed in stainless-steel containers filled with Ar gas for growing single crystals.
The containers were then heated at 950$^\circ$C for 10 h and cooled to 550$^\circ$C at 2$^\circ$C/h.

To characterize the crystalline quality of the compounds, back-reflection Laue X-ray measurements were performed using an s-Laue instrument (PULSTEC). 
The lattice constants $c$ were determined from X-ray diffraction measurements (XRD) using a SmartLab diffractometer (Rigaku) with Cu k$\alpha_1$ radiation. 
The chemical compositions were examined using energy-dispersive X-ray spectroscopy (EDX) on an AZtec One instrument (Oxford). 
The Seebeck coefficients and electrical resistivities above $T$ = 300 K were examined along the a-axis direction using an RZ-2001i instrument (Ozawa). 
The electrical resistivities below $T$ = 300 K were evaluated using a homemade measurement system.

\begin{figure}
\begin{center}
\includegraphics[width=1\columnwidth]{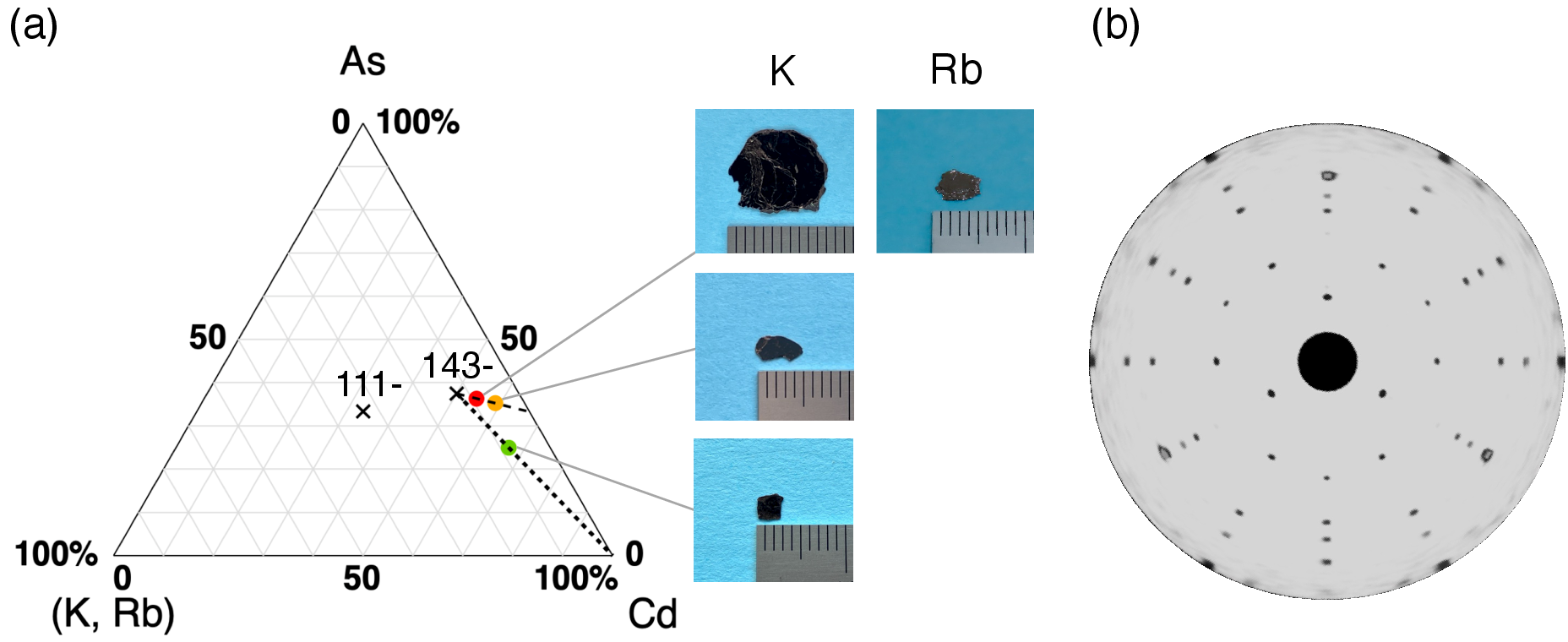}
\caption{\label{phasediagram}(Color online)(a) Ternary phase diagram of the (K or Rb) - Cd - As system with photographs of grown single crystals.  
The closed circles depict the starting compositions of the single crystal growth.
The dotted and long dashed lines demonstrate the composition lines of the Cd and Cd$_2$As flux methods, respectively. 
(b) Back-reflection Laue photograph of a grown KCd$_4$As$_3$ single crystal.
}
\end{center}
\end{figure}

First, we grew KCd$_4$As$_3$ single crystals using the Cd flux method, as previously reported for growing NaCd$_4$As$_3$ single crystals. \cite{NaCd4As3_topo, NaCd4As3_Hall, NaCd4As3_2025}
Single crystals with a typical size of 3 $\times$ 3 $\times$ 0.2 mm$^3$ were successfully grown from the starting materials KAs, Cd, and As at a ration of 1 : 8 : 2 (Fig. \ref{phasediagram}(a)).
However, KCdAs was precipitated as an extra phase inhibiting crystal growth.
To avoid such precipitation, a Cd$_2$As-flux located opposite the KCdAs phase in the ternary phase diagram was applied. 
In addition, the flux ratio was reduced as  the stoichiometric composition was approached to enlarge the grown crystals. 
 Starting compositions of KAs:Cd$_2$As = 1:5 and 1:3, corresponding to KCd$_4$As$_3$:Cd$_2$As = 1:3 and 1:1, respectively, yielded single crystals of KCd$_4$As$_3$ without the KCdAs phase. 
The grown single crystals were sized 5 $\times$ 3 $\times$ 0.2 mm$^3$ (KCd$_4$As$_3$:Cd$_2$As = 1:3) 
and 10 $\times$ 9 $\times$ 0.2 mm (KCd$_4$As$_3$:Cd$_2$As = 1:1). 
Larger single crystals were obtained from a starting composition near the stoichiometry. 

Having successfully grown KCd$_4$As$_3$ single crystals, we grew RbCd$_4$As$_3$ single crystals under a Cd$_2$As flux.
Large-size RbCd$_4$As$_3$ single crystals (5 $\times$ 3 $\times$ 0.3 mm$^3$) were successfully grown from a starting composition of RbAs:Cd$_2$As = 1:3 (RbCd$_4$As$_3$:Cd$_2$As = 1:1).

EDX measurements on the single KCd$_4$As$_3$ and RbCd$_4$As$_3$ crystals confirmed stoichiometric compositions of K : Cd : As = 1.03 (1) : 4.01 (1) : 2.96 (1) and Rb : Cd : As = 1.03 (1) : 4.00 (1) : 2.97 (1) for KCd$_4$As$_3$ and RbCd$_4$As$_3$, respectively. 
From the XRD measurements, the lattice constants of KCd$_4$As$_3$ and RbCd$_4$As$_3$ were determined as $c$ = 36.0690(3) {\AA} and 37.1040(4) {\AA} respectively, consistent with previous reports\cite{RbCd4As3_Ono,143_He}.
The back-reflection Laue X-ray photograph demonstrated a three-fold rotational symmetry, 
confirming the wide facet of the specimen as the $c$ plane (Fig. \ref{phasediagram}(b)).

\begin{figure}
\begin{center}
\includegraphics[width=0.7\columnwidth]{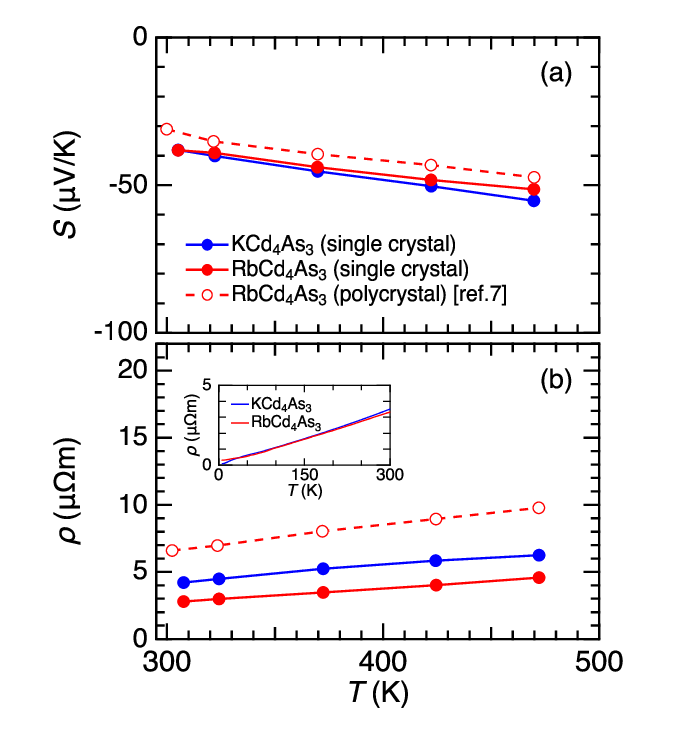}
\caption{\label{ozawa}(Color online)
Temperature dependences of (a) the Seebeck coefficient ($S$) and (b) the electrical resistivity ($\rho$) of single crystal KCd$_4$As$_3$ and RbCd$_4$As$_3$ 
measured along the $a$-axis.
Data of polycrystalline RbCd$_4$As$_3$ (open circles) are plotted as a reference\cite{RbCd4As3_Ono}.
The inset in Fig.  \ref{ozawa}(b) shows the electrical resistivity of single crystal KCd$_4$As$_3$ and RbCd$_4$As$_3$ below $T$ = 300 K.

}
\end{center}
\end{figure}

The Seebeck coefficients of the KCd$_4$As$_3$ and RbCd$_4$As$_3$ single crystals were negative, indicating that both compounds are n-type conductors (Fig. \ref{ozawa}(a)). 
The magnitudes of the Seebeck coefficients including that of polycrystalline RbCd$_4$As$_3$ were equivalent, indicating independent carrier density on the alkali atoms.

The electrical resistivities along the $a$-axis were linear functions of temperature like a metal (Fig. \ref{ozawa}(b)).
The values of RRR, defined as $\rho_a$(300 K)/$\rho_a$(5 K), were 37 and 11 for KCd$_4$As$_3$ and RbCd$_4$As$_3$ single crystals, respectively. 
The RbCd$_4$As$_3$ single crystal exhibited a smaller electrical resistivity than the polycrystalline sample.
The electrical resistivity in the polycrystalline sample ($\rho_p$) should be approximately 3/2 of resistivity along $ab$-plane in single crystal ($\rho_{ab}$); that is, $\rho_p$ = 3/2$\rho_{ab}$
because the high resistivity along the $c$-axis restricts the cross-sectional area of electrical conduction in polycrystalline RbCd$_4$As$_3$ to 2/3.
The anisotropic conduction was supported by the band calculation where 
the calculated effective masses along the $ab$-plane were 5-7 times smaller than those along the $c$-axis\cite{RbCd4As3_Ono}. 
On the other hand, they were approximately isotropic in the $ab$-plane.
In fact, the electrical resistivity along the $a$-axis in single crystal RbCd$_4$As$_3$ ($\rho_a$) was approximately 2/3$\rho_p$ (Fig. \ref{ozawa}(b)). 
Slight downward deviation might be due to the suppression of charge carrier scattering at the grain boundaries.

Meanwhile, KCd$_4$As$_3$ single crystal exhibited a higher electrical resistivity than the RbCd$_4$As$_3$ single crystal, possibly attributable to the heavier effective mass of KCd$_4$As$_3$.
The effective mass of NaCd$_4$As$_3$ \cite{NaCd4As3_topo,NaCd4As3_Hall} is approximately twice that of RbCd$_4$As$_3$, 
suggesting that the effective mass increases with decreasing mass of alkali metal atom. 

At $T$ = 472 K, the power factors of the KCd$_4$As$_3$ and RbCd$_4$As$_3$ single crystals were calculated as $PF$ = 0.49 and 0.58 mW/mK$^2$, respectively.
The $PF$ of single crystal RbCd$_4$As$_3$ was approximately three times that of polycrystalline RbCd$_4$As$_3$ ($PF$ = 0.23 mW/mK$^2$), owing to the lower electrical resistivity of single crystal RbCd$_4$As$_3$.


In summary, large-size single crystals of KCd$_4$As$_3$ and RbCd$_4$As$_3$ were successfully grown using the Cd$_2$As flux method. 
Single crystal KCd$_4$As$_3$ and RbCd$_4$As$_3$ exhibited negative Seebeck coefficients independently of their alkali atoms, demonstrating their n-type conductivity.
Their electrical resistivities increased linearly with temperature, consistent with those of polycrystalline samples.
RbCd$_4$As$_3$ delivered the maximum $PF$ (0.58 mW/mK$^2$) at $T$ = 472 K, demonstrating its promising potential as a high-performance thermoelectric material.

\begin{acknowledgment}
The author thanks K. Kihou and S. Ishida for valuable discussions. 
This work was supported by Mirai (JPMJMI19A1) and SPRING (JPMJSP2123) from the Japan Science and Technology Agency.
\end{acknowledgment}




\begin{thebibliography}{9}
\bibitem{NaZn4As3_aichi}A. Yamashita, K. Kihou, H. Kunioka, H. Nishiate, A. Yamamoto, Y. Goto, Y. Mizuguchi, T. Iida, Y. Takano, and C. H. Lee, J. Solid State. Chem. \textbf{291}, 121588 (2020).

\bibitem{RbZn4As3_Ono}K. Ono, K. Kihou, H. Usui, K. Kuroki, Y. Goto, and C. H. Lee, ACS Omega \textbf{8}, 42900 (2023).


\bibitem{RbZn4P3calc_2022}Sangeeta, R. Kumar, and M. Singh, J. Mater. Sci. \textbf{57}, 10691 (2022).

\bibitem{CsZn4As3calc_2024}Y. Xue, C. Lin, B. Huang, H. He, D. Huang, and C. Persson, J. Solid. State. Chem. \textbf{339}, 124976 (2024).

\bibitem{CsMn4As3calc_2025}S. S. Sahoo, H.-J. Koo, M.-H. Whangbo, G. Vaitheeswaran, and V. Kanchana, J. Phys.: Condens. Matter \textbf{37}, 205801 (2025).

\bibitem{CsCd4As3calc_2026}A. Jabar, Z. Fadil, S. Idrissi, L. Bahmad, M. N. Khan, C. J. Raorane, and S.-C. Kim, J. Mol. Graph. Model. \textbf{146}, 109433 (2026).


\bibitem{RbCd4As3_Ono}K. Ono, Y. Goto, H. Usui, M. Murata, S. Tsutsui, K. Kuroki, Y. Kamihara, and C. H. Lee, Acta Mater. \textbf{307}, 121968 (2026).

\bibitem{NaCd4As3_topo}Y.-Y. Wang, C. Zhong, M. Li, W.-L. Zhu, W.-J. Hou, W.-H. Song, Q.-X. Dong, Y.-F. Huang, S. Zhang, Z.-A. Ren, S. Wang, and G.-F. Chen, Phys. Rev. B \textbf{102}, 115122 (2020).

\bibitem{NaCd4As3_Hall}H. Yang, Y. Zhou, J. Wang, Y. Zhou, Y. Han, J. Wang, Y. Li, X. Zhang, and Z. Yang, Appl. Phys. Lett. \textbf{125}, 203101 (2024).

\bibitem{NaCd4As3_2025}T. R. Kafle, Y. Zhang, Y. Y. Wang, X. Shi, N. Li, R. Sapkota, J. Thurston, W. You, S. Gao, Q. Dong, K. Rossnagel, G.-F. Chen, J. Freericks, H. C. Kapteyn, and M. M. Murnane, Struct. Dyn. \textbf{12}, 014501 (2025).

\bibitem{NaZn4As3_topo}Q.-X. Dong, B.-B. Ruan, Y.-F. Huang, Y.-Y. Wang, L.-B. Zhang, J.-L. Bai, Q.-Y. Liu, J.-W. Cheng, Z.-A. Ren, G.-F. Chen, Chin. Phys. B \textbf{32}, 066501 (2023).

\bibitem{RbCd4As3_calc}N. Sachdeva, R. Kumar, and M. Singh, AIP Conf. Proc. \textbf{3044}, 050006 (2024).



\bibitem{Kihou_KFe2As2}K. Kihou, T. Saito, S. Ishida, M. Nakajima, Y. Tomioka, H. Fukazawa, Y. Kohori, T. Ito, S. Uchida, A. Iyo, C. H. Lee, and H. Eisaki, J. Phys. Soc. Jpn. \textbf{79}, 124713 (2010).


\bibitem{143_He}H. He, C. Tyson, and S. Bobev, Inorg. Chem. \textbf{50}, 8375 (2011).


\end{thebibliography}
\end{document}